\documentclass[twocolumn,showpacs,preprintnumbers,amsmath,amssymb,superscriptaddress]{revtex4}
\pdfoutput=1

\usepackage{dcolumn}
\usepackage{bm}

\usepackage{amsfonts}
\usepackage{amssymb}
\usepackage{graphicx}
\usepackage{latexsym}
\usepackage{amsmath}
\def\ee{\end{eqnarray}}

\newcommand{\be}{\begin{eqnarray}}
\newcommand{\en}{\end{eqnarray}}
\newcommand{\bea}[1]{\left(\begin{array}{#1}}
\newcommand{\ena}{\end{array}\right)}

\newcommand{\tmop}[1]{\ensuremath{\operatorname{#1}}}
\begin{document}

\preprint{}

\title{A Simple Explanation for DAMA with Moderate Channeling}

\author{Brian Feldstein}
\affiliation{Physics Department, Boston University, Boston, MA,
02215}
\author{A. Liam Fitzpatrick}
\affiliation{Physics Department, Boston University, Boston, MA,
02215}
\author{Emanuel Katz}
\affiliation{Physics Department, Boston University, Boston, MA,
02215}
\author{Brock Tweedie}
\affiliation{Department of Physics and Astronomy, Johns Hopkins
University, Baltimore, MD 21218}

\date{\today}

\begin{abstract}
We consider the possibility that the DAMA signal arises from channeled events 
in simple models where the dark matter interaction with nuclei is
suppressed at small momenta.  As with the standard WIMP, these 
models have two parameters (the dark matter mass and the 
size of the cross-section), without the need to introduce 
an additional energy
threshold type of parameter.  We find that they can be consistent
with channeling fractions as low as about $\sim 15$\%, so long as at least
$\sim$70\% of the nuclear recoil energy for channeled events is deposited 
electronically.
Given that there are reasons not to expect very large channeling fractions,
these scenarios make the channeling explanation of DAMA
much more compelling.
\end{abstract}


\maketitle

\section{Introduction}
\label{sec:intro}
  The annual modulation signal seen by the Dark Matter (DAMA) collaboration,
  now significant at the 8$\sigma$ level \cite{DAMALIBRA,DAMA}, 
provides an intriguing puzzle. If
  one assumes that this signature is due to a standard weakly interacting
  massive particle (WIMP) elastically scattering off nuclei, then it follows
  that other direct detection experiments should also have seen large numbers
  of signal events. This is, however, not the case.
  As a result, one is left to explore dark matter
  scenarios going beyond the standard elastic WIMP paradigm. 
   Indeed, it is not
  currently possible to rule out a dark matter explanation for DAMA in a model
  independent way. The list of known possibilities still consistent with all
  available data, however, is in fact rather limited.
  
 One scenario which has received a great deal of attention is that of
 inelastic dark matter \cite{Neal1}. In this picture, one assumes that in each scattering
 event the dark matter particle makes a transition to a higher energy state,
 separated from the original by a splitting $\delta$. This splitting
 effectively removes scattering events at sufficiently small nuclear recoil
 energies, where standard WIMP event rates become large and cause the
 greatest tension with data. Another possibility is that the events at
 small recoil energies become suppressed by a dynamical form 
factor \cite{OurPaper,Chang:2009yt}.
 Yet another option is that the scattering proceeds through a 
resonance \cite{Fox}.
  
In all of the above cases, however, great care must be taken to
ensure that the DAMA signal is present, without resulting in an
over-abundance of events at other experiments. \ In inelastic dark matter,
a splitting is added solely to remove events at experiments
with light nuclei.  This splitting cannot be too small, 
lest the low energy suppression
be insufficient, but it also cannot be too large, or else the DAMA signal
disappears.  Hence, for any given dark matter mass, $\delta$ must fall in 
a narrow range, and additional model building is required to motivate
the coincidence associated with its size. 

In the dynamical form factor scenario, careful model building must be
undertaken in order to ensure that the required low energy suppression is
sufficiently rapid.  In explicit models constructed so far,
here too there is a parameter (analogous to $\delta$) that
must be carefully 
chosen to make the DAMA signal consistent with 
the null experiments.  

While such model building avenues are currently still viable for
explaining DAMA, perhaps a nicer resolution would be obtained if there were
a reason found for DAMA simply being the most sensitive direct detection 
experiment at low recoil energies, where WIMP signals tend to be peaked. 
In fact, ``channeling'' has already been proposed as an explanation of this 
type \cite{Channel1, zurek,fairbairnschwetz,SpencerAaronNeal,lisapoland}.  
Most direct detection experiments, including DAMA, do not measure the complete recoil energy of a
nucleus in a scattering event, but only some fraction of it.
In particular, DAMA measures only the fraction deposited electromagnetically.
The complete
recoil energy is then inferred through multiplication by an experimentally
determined ``quenching factor''. 
DAMA, modeling their quenching factor to be the same for all
recoiling Iodine nuclei, determines it to have a value of 0.09. 
%\ These quenching factors are often assumed
%to be independent of energy, and are not always directly tested at all of the
%energies relevant to an experiment. \ In the case of DAMA, only
%electromagnetic energy in a scattering event is detected, leading to a
%quenching factor for Iodine recoils of approximately $.09$. \ 
%%%{\bf {I want to say something more specific here about the nature of the experimental
%uncertainty in DAMA's quenching factor, but I dont understand the details well
%enough at the moment}.  
The idea behind channeling is that, due to the crystal structure
of the detector, some fraction of low energy events may actually have a
quenching factor closer to one. \ This is the expectation for
nuclear recoils which travel sufficiently 
%%%
along an axis of the crystal.
%far from the atoms of the crystal
%along one of its axes. \ 
For such ``channeled'' events, 
a nuclear recoil which was thought to have had an energy of, say $\sim 20 \tmop{keV}$, actually
had an energy closer to $\sim 2 \tmop{keV}$. 

%Since WIMP event rates are
%expected to rise sharply towards low energies, to the extent that channeling
%is occuring, this would imply that the detector is actually a much more
%sensitive instrument than was originally thought. \ This could then lead to a
%beautiful explanation for DAMA's rather mysterious experimental result.

In this letter, we will reconsider the possibility of explaining the
DAMA data through the assumption that some fraction of recoiling nuclei
are being channeled.  Since we believe that current theoretical estimates for
the amount of channeling are most likely not reliable, our
goal will be to try to explain the DAMA data using as little channeling as
possible. 
%It is also unlikely that an ion recoiling from a 
%a lattice site (after its collision with the DM particle), can 
%become channeled without losing some fraction of its energy to the 
%the lattice.  We are therefore also interested in viable scenarios were
%the quenching factor need not be extremely close to one.
%
%Taking into account
%some level of uncertainty 
%only modest variations
%in the
%dark matter halo velocity distribution, 
We will show that unlike the case of a 
standard WIMP, the simplest momentum dependent couplings can 
allow for an appreciable decrease in the amount of
channeling required to explain the data, to as little as $\sim 10$\%.
In addition, for channeling fractions of at least $\sim$ 15\%, these couplings
can allow for channeled events to have a quenching factor
as small as $\sim 70$\%
%a more mild quenching factor 
and still satisfy all constraints.
%while remaining largely insensitive to possible energy dependence
%in the channeling fraction.    

No complicated model building is needed
to arrange for the required momentum dependence.  
For example, 
it is sufficient
to have a neutral dark matter particle, $X$, 
%with a charge-radius 
whose leading interaction 
with a new GeV mass gauge boson $A_\mu$ 
therefore becomes the lowest dimension gauge-invariant interaction
${\cal L} = \frac{i}{\Lambda^2} F_{\mu\nu} \partial^\mu X^* \partial^\nu X.$
Kinetic mixing between $A_\mu$ and hypercharge then couples
$X$ to nuclei in a momentum dependent fashion.  The mass of the 
gauge boson and the scale $\Lambda\sim$GeV are both absorbed into the 
overall scattering cross-section, and so constitute a single parameter.

The scenario discussed here therefore offers a solution to the DAMA
puzzle
that does not require extra energy threshold parameters, and is 
viable with more conservative and realistic versions of channeling.
Finally, in the event that heavy element experiments such as CRESST 
\cite{CRESST2, CRESST},
 XENON \cite{XENON10}, and KIMS \cite{KIMS} 
fail to find evidence for dark matter in the near future, 
this scenario may remain as an attractive viable option.

%The outline of this paper is as follows: \ In section ?? we will
%briefly review the ingredients of direct detection event rate calculations. \
%In section ??, we will review the basics of the physics of channeling, as well
%as note some uncertainties in the associated theoretical predictions. \
%Section ?? will discuss some examples of simple interactions leading to the
%low energy suppressions of interest, while allowed regions of parameter space
%will be presented in section ?. \ We will conclude in section ?.

% \ A similar situation exists for the form factor scenario. \
%  (does it? \ how to characterize this for a general form factor?)
%  
%  
%  
%  
%  
%  dama sees blah blah blah
%  
%  everything sucks a bit...
%  
%  channeling is intriguing possibility, doesn't quite work
%  
%  mention inelastic + channeling
%  
%  but would be nice to have an explanation that doesn't depend on coincidental
%  parameter, no new scales

\section{Review}

The event rate per unit detector mass per unit recoil energy at a dark matter
direct detection experiment is given by (see e.g.
 \cite{lewinsmith,lewinsmithlong})
\be
\frac{d R}{d E_R} = N_T \frac{\rho_{\rm DM}}{m_{\rm DM}}\int_{v_{min}} d^3v f(\overrightarrow{v}) v \frac{d \sigma}{d
  E_R}.
\label{eq:drder}
\ee
In this expression, $N_T$ is the number of target nuclei per unit mass,
$\rho_{\chi}$ is the local dark matter density and $m_{DM}$ is the mass of
the dark matter particle.  The integration is over the distribution of dark
matter velocities $f(\overrightarrow{v})$ relative to the Earth; the lower
bound $v_{min}$ is given by the minimum velocity a dark matter particle must
have in order to cause a nuclear recoil with energy $E_R$.  In particular, we
have $v_{min} = \frac{1}{\mu}\sqrt{\frac{m_N E_R}{2}}$, where $m_N$ is the
mass of the recoiling nucleus, and $\mu$ is the dark matter-nucleus reduced
mass. 
%{\bf not quite right - include earth's velocity $v_e = 220 $ 
%km/s (btw, note
%that $f(v)$ in integral above is in earth rest frame)}.

The cross section for a dark matter particle to scatter off of a nucleus with charge
Z and atomic number A is given by
\be
\frac{d \sigma}{d E_R} = \frac{m_N}{2 v^2} \frac{\sigma_p}{\mu_n^2} \frac{(f_p
Z + f_n (A-Z))^2}{f_p^2} F_N^2(E_R) F_{\rm DM}^2(E_R). 
\label{eq:xsec}
\ee
%{\bf put in DM form factor in this formula!}
Here $\mu_n$ is the dark matter-nucleon reduced mass, 
$f_n$ and $f_p$ are
the relative coupling strengths to protons and neutrons, and $F_N$ is the
nuclear form factor. 
 $F_{\rm DM}$ is a possible form factor coming from the dark
matter sector, which 
is equal to $1$ for a standard WIMP,
and in general is normalized to be 1 at momentum transfer $q = \sqrt{2m_{\rm N} E_R}=$
30 MeV.  $\sigma_p$ is an overall constant that would
be the dark
matter-proton cross section in the case where $F_{\rm DM}\equiv 1$.
 For simplicity we will take $f_n = 0$ in this letter,
corresponding, for example, to scattering taking place through a new $U(1)$
gauge symmetry which mixes with hypercharge.

The form of the dark matter velocity distribution is significantly uncertain.
It is generally considered that a Boltzmann distribution $\propto
e^{-v^2/\bar{v}^2}$ in the galactic rest frame is a reasonable starting
point.  We will consider the cases $\bar{v} =170$km/s, 220km/s, and 270km/s, 
and
take the distributions to be cut off at an escape velocity $v_{\rm esc}$ 
whose value we shall take to be 500km/s \cite{vesc}.  

We will parameterize the velocity $v_e$ of the Earth's galactic
motion, as well as the nuclear form factors, as in \cite{OurPaper}. 

\section{Channeling}

Channeling of fast-moving ions through crystal lattices is a well-established physical phenomenon with an extensive literature.  For a relatively recent review, see \cite{Cohen:2004}.  The basic theory, developed by Lindhard in the sixties \cite{Lindhard}, details under what conditions an ion will become guided moving along a particular axis or within a particular plane by the cumulative effects of soft collisions with the lattice atoms.  In this theory, a channeled ion never gets close enough to an individual atom to lose a significant amount of energy from elastic scattering.  Typically, such an ion will instead lose most of its energy by electronically exciting the atoms it passes.

If a WIMP strikes a nucleus in the DAMA detector, it is ejected from its
position in the NaI lattice (dragging along most of its electrons \footnote{In
  an intuitive semiclassical picture due to Migdal \cite{Migdal:1941}, electrons orbiting
  at velocities greater than the nucleus's recoil velocity will be largely
  unaffected by the scattering.  
%At DAMA, the maximum possible recoil
%  velocity (from an encounter with a very heavy WIMP) is twice the escape
 % velocity of the galaxy, or roughly $4\times10^{-3}c$.  
For the recoil velocities relevant at DAMA, this picture applies for
almost all of the electrons in both Na and I atoms, and thus most of the
electrons will remain in their original orbitals.}) and becomes a fast-moving ion.  As pointed out by Drobyshevski \cite{Channel1}, this recoiling ion could in principle become channeled, making DAMA sensitive to much lower energies than originally expected.

There are two crucial questions here:  1) What is the probability for a recoiling
ion to become channeled at $O(1\sim10)$ keV energies in NaI?, and 2) On
average, what fraction of the recoil energy of such channeled ions is lost
electronically?  In \cite{channel2}, the DAMA collaboration made a first attempt at
answering these questions, and using Lindhard's theory, estimated the
effects to be significant.  For Na and I ions at 3 keV, they found 
channeling
probabilities of roughly 30\%, and took quenching factors of almost 100\%.

However, rigorous experimental and theoretical studies of this regime in NaI
are lacking, and the full physical situation remains far from clear.  At DAMA
energies, calculation of the critical trajectories for channeling falls
outside the realm of applicability of the Lindhard theory, as the details of
the collective lattice potential become important.  Other novel
complications appear, as well. 
% To date, channeling has never been observed in nuclear recoil, but only in
% externally-injected ion beams.  
%There may be a good reason for this:  
For example, a recoiling nucleus necessarily originates {\it at} a lattice
site, and thus if it is aimed towards a crystal axis or plane, its trajectory  will tend to lie in the direction of its neighbors.  It might then
immediately undergo scattering away from the intended channel
\footnote{Within the context of the Lindhard theory, an ion will only end up
  channeled if it does not penetrate too close to a guiding string or plane
  of lattice atoms.  This criterion is manifestly violated if the starting
  position of the ion is already within this string or plane.}.  Even if a
recoiling ion manages to find itself channeled after a few atomic scatterings
with modest energy loss, it is not obvious that the subsequent electronic
energy losses will dominate for such a low-energy channeled ion (see,
e.g. \cite{Kumakhov:1981}).  Above all else, it is clear that an experimental study of the channeling of
recoiling nuclei in NaI would be extremely valuable.

In the rest of this paper, we will take an agnostic viewpoint towards these
issues, and simply parametrize the possibility of a population of recoiling
ions depositing an anomalously large fraction of its energy electronically.
In the end, even a somewhat large non-gaussian tail of the event-by-event
quenching factor can lead to the effects predicted by Drobyshevski, and
channeling need not even be the primary cause of this.  Regardless of its
origin, we will continue to call this hypothetical effect ``channeling'' for
continuity.

Should an effect like this occur at the level claimed by DAMA, the
implications for their observed annual modulation at $\sim3$ keV would be
quite dramatic.  For light WIMPs with standard SI interactions, the absolute
and modulating spectra peak sharply at low energies, and DAMA becomes capable
of directly sampling part of this otherwise unobservable region at $E_R \sim3$
keV.  This causes a portion of the DAMA allowed region in the
($m$,$\sigma_N$) plane to edge past constraints from CDMS 
\cite{CDMS,CDMS2,CDMSFIVE} and XENON
\cite{zurek,fairbairnschwetz}.
%But the story is not quite so simple for the light standard WIMP.  First, the
%DAMA modulation spectrum [XXX] displays a marginally significant dip at the
%lowest energy bin.  This is to be contrasted with the expected steep rise
%towards low energy.  Second, the experiment CoGeNT [XXX] has probed to lower
%energies than any prior experiment, $0.5$ keVee (or nominally about 3 keVnr,
%unfolding the quenching factor).  It sees no evidence for a light WIMP.  If
%channeling is similarly important in the diamond-like lattice structure of
%CoGeNT's Ge crystal detector (versus salt-like for NaI), then it
%unambiguously becomes more sensitive than DAMA, excluding the light WIMP
%option.
On the other hand, as we will see in the next section, even a modest
relaxation of either the assumed channeling fraction or its effective
quenching factor will move all of the allowed region back into conflict with
these experiments.

A simple, more robust alternative is a dark matter particle
with momentum-depedent interactions 
due to form factors or other explicit $E_R$-dependence in
the scattering matrix elements.
%{\bf Brock, what citation had you intended to put here?}  
%Then it becomes natural for the first DAMA modulation bin to measure a lower
%value, and constraints from CoGeNT can essentially disappear, even
%accounting for Ge channeling.  
As we will see below, even modest amounts of channeling could make simple
form factors in the dark sector a viable explanation for all of the present data.

%To Do:
%\begin{itemize}
%\item Look around for some data on channeling of recoiling nuclei.  Has this really ``never been observed''?
%\item Try to get a better estimate for electronic stopping power of channeled ions.  Nuclear stopping powers are typically about $10^{-14} {\rm eV cm}^2/{\rm (atomic layer)}$, or $\sim30$ eV/nm.
%\end{itemize}

%Questions for BU:
%\begin{itemize}
%\item Is the unchanneled DAMA allowed sliver, shown in the CoGeNT paper, still alive?  If so, how can we put any constraints on the channeling parameters?
%\item Do you guys understand planar channeling?
%\item How important is I scattering vs Na?
%\end{itemize}

\begin{figure}[t!]
\begin{center}
\includegraphics[width=0.5\textwidth]{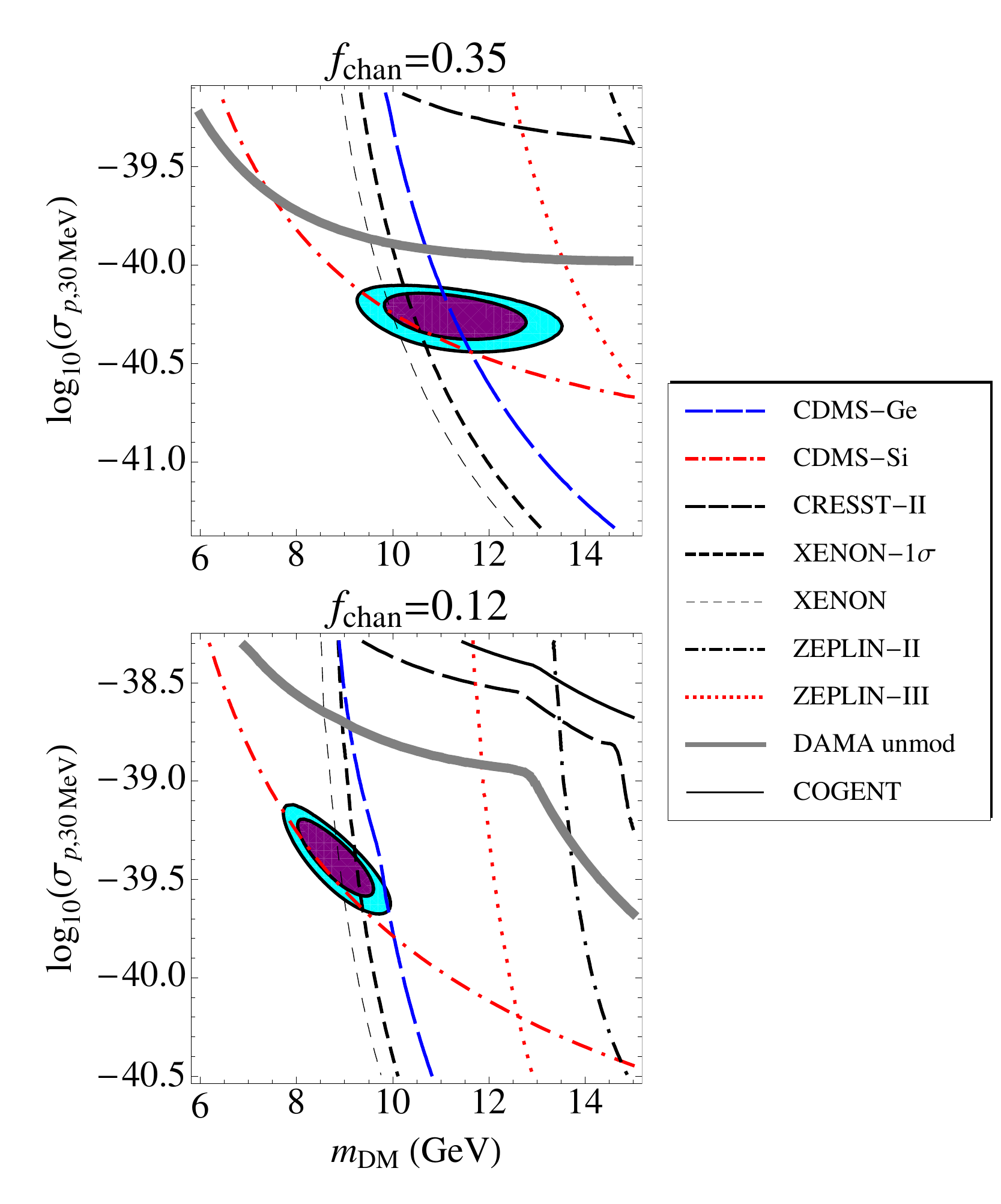}
\caption{Constraint plots for channeled events, with an energy-independent
channeling fraction. The bottom plot shows constraints on models with
a momentum-dependent $F_{\rm DM}(q) \propto q^2$ coupling, and
the top assumes a standard (i.e. momentum-independent) coupling. 
DAMA 90\% and 99\% contours are shown. The $f_{\rm chan}$'s above
are chosen to be the minimum values consistent with all constraints.
Obviously, larger values of $f_{\rm chan}$ leave more parameter space open.}
\label{fig:banana}
\end{center}
\end{figure}

\begin{figure}[t!]
\begin{center}
\includegraphics[width=0.48\textwidth]{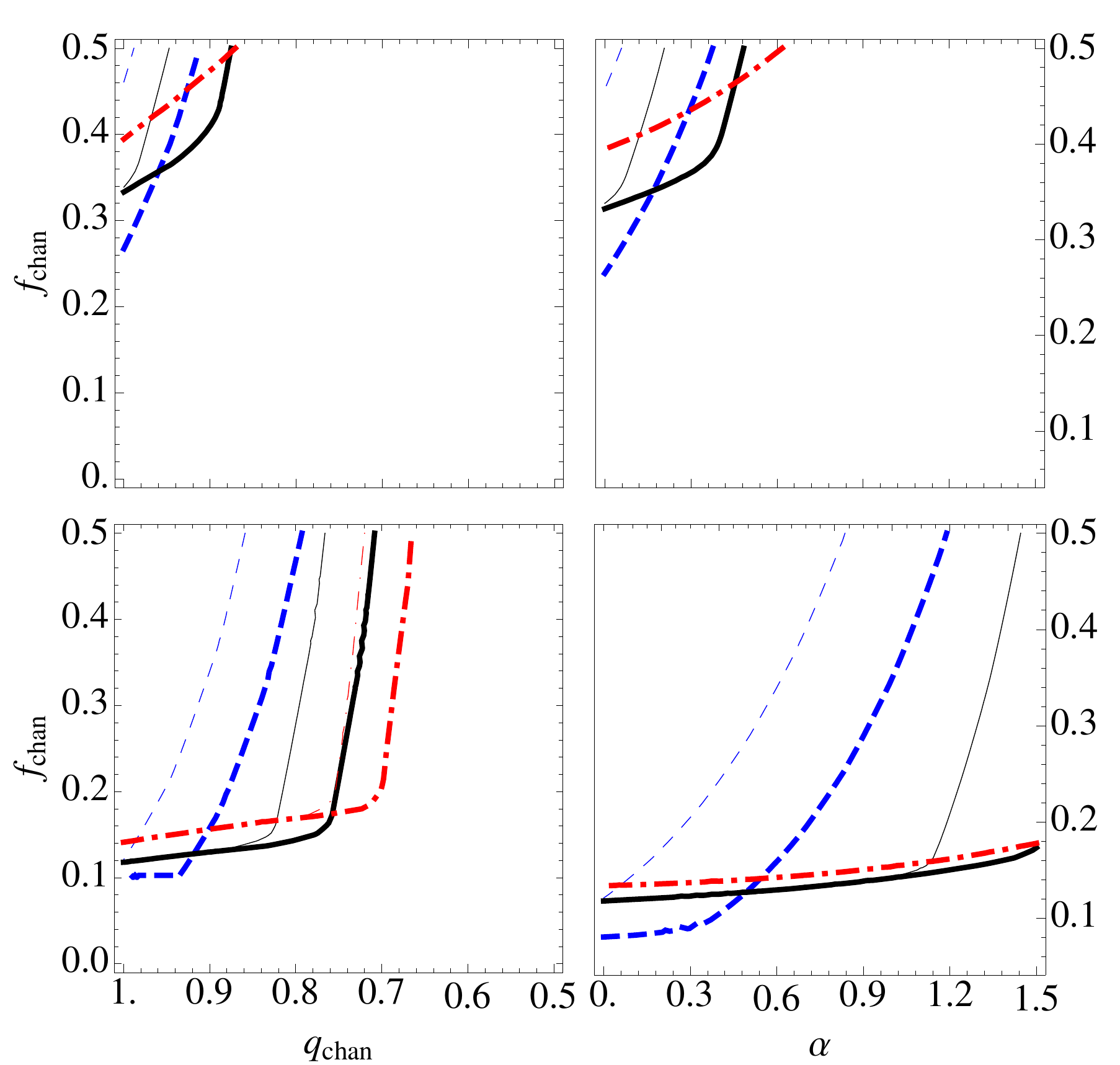}
\caption{ Minimum channeling fractions allowed at $E_R = 3$ keV depending
on the channeled quenching factor $q_{\rm chan}$ and the channeling fraction
energy-dependence, $f(E_R) \propto E_R^{-\alpha}$.  Constraints are shown
for $v_0=170,220,270$km/s in blue dashed, black solid, red dot-dashed,
respectively,
and $q_{\rm XENON} =$15.5\%(14.3\%)
in thin (thick) lines.  The top panels are for a momentum-independent 
coupling and
the bottom are for $F_{\rm DM} \propto q^2$. Left (right) has $\alpha=0$
($q_{\rm chan}=1.$). The 
$v_0=$270km/s constraint in the lower right plot is identical for
the two values of $q_{\rm XENON}$ we consider.}
\label{fig:qandpow}
\end{center}
\end{figure}

\section{Required Channeling Fraction and Shape}

If the DAMA signal is due to channeled events, then 
the constraints from many other direct detection experiments
can be avoided because most of their energy range would correspond
to dark matter events with velocity above the local escape velocity.  
It is useful to convert
the energy range of experiments into a range in momentum transfers
$q$, with $q_{\rm min}$ the lowest momentum transfer probed.  
Then a given experimental constraint is completely evaded for dark
matter masses below some 
critical value, $m_{\rm DM}^{-1} = 2(v_{\rm esc} + v_{\rm e})
/q_{\rm min} - m_{\rm N}^{-1}$.  In particular, it is possible to choose
$m_{\rm DM}$ so that all of the CDMS-Ge and CRESST-W range in $q$ is above
the escape velocity, while all of the DAMA signal region is below it. 
The most significant constraints then become CDMS-Si and, depending
on its quenching factor, XENON.  In addition, it is
important that values of  $m_{\rm DM}$ below the critical value still
give a decent fit to the DAMA spectrum.  We will see that model-dependent
assumptions can affect the quality of the fit significantly, allowing
or disfavoring smaller dark matter masses.

Because the critical mass for a given experiment depends on $q_{\rm min}$,
and event rates tend to rise sharply at decreasing $q$,
constraints are very sensitive to even modest uncertainties in
quenching factors, which relate the experimental light yield to the 
actual recoil energy.  XENON in particular has measured its quenching factor
to be 15.5 \% $\pm$ 1.2\% (weighted average over \cite{aprile,chepel,sorensen},
but see also \cite{Manzur:2009hp}, which indicates that the experiments
may be underestimating their uncertainties).  We will
therefore present constraints from XENON using both $q_{\rm XENON}=15.5$\%
and the lower $1 \sigma$ value 14.3\% (``XENON-$1\sigma$'' for short).

There are a number of directions in which the parameter space opens
up if one takes into account uncertainties or variations in models.  
The quality of the fit to DAMA depends on the shape of the predicted
spectrum, and including an additional dark matter form factor
$F_{\rm DM}(q) \sim q^2$ allows a range of smaller dark matter masses.
%Such a form factor naturally arises in models of dark matter;  if the dark
%matter is a neutral scalar composite field whose constituents are charged
%under a dark force that mixes with the photon, then the lowest dimension
%operator allowed generates exactly this form factor \cite{OurPaper}.  

It is important to take into account 
that the masses favored by DAMA are also sensitive
to the energy-dependence of the channeling fraction.  Most if not all
studies of channeled events assume a channeling fraction of the form
$f_{\rm I}(E_R) = {e^{-E_R/40} \over 1+0.65 E_R}$ or something very close
\cite{zurek,JMR,SpencerAaronNeal}, which over the
range $2 {\rm keV} < E_R < 6$ keV relevant at DAMA is approximately
$\propto E_R^{-0.75}$.  However, as we have discussed, this estimate
is based on a relatively simple analytic approximation due to Lindhard, 
and for example
if the channeling fraction is saturated at the relevant low recoil energies,
then the profile could instead be approximately constant
\footnote{An amusing possibility is that the turn-over in the DAMA 
spectrum at low energies could even be due to a turn-over
in the channeling fraction, if channeling became less
effective below $E_{R}\sim$3keV.  }.  Such a profile
would favor somewhat smaller masses, which in turn push more of the
null experiments' energy ranges above the escape velocity.  Figure
\ref{fig:banana} shows the 90\% constraints on a WIMP and the
90\%, 99\% DAMA fit contours if the channeling
fraction is energy-independent. A channeling fraction of
$\approx$35\% is the minimum value required to have some region of parameter space allowed by all constraints.  The situation is even better with an
additional dark matter form factor $F_{\rm DM} \propto q^2$.  In this
case, much of the DAMA favored masses are below the critical value $m_{\rm DM}$ 
where
{\it no} events are predicted at XENON or CDMS-Ge, and the minimum
channeling fraction allowed becomes $\approx $12\%.
In all of the allowed parameter space, iodine scattering always dominates
over sodium scattering at DAMA \footnote{
We have checked that even if COGENT and KIMS have channeling fractions
comparable to those at DAMA, they are still subdominant to other null
experiment constaints.}.

  Our analysis assumptions
are as in \cite{OurPaper} except for XENON, where we vary the quenching factor,
CRESST-II, where we use only the 2007 run, and CDMS-Si, where we have
taken an effective exposure of 12 kg days (from March 25 - Aug 8) and no events.  
Also, 
one of the ten potential XENON events was identified by that collaboration 
as resulting from instrumental error, and we do not include it in calculating
our constraints.

\begin{table}[t!]
\vspace{0.5cm}
\begin{tabular}{|l|l|l|l|}
    \hline
$n$ & flat profile & Lindhard profile \\
\hline
\hline
\multicolumn{3}{|c|}{ $v_0=170$ km/s} \\
\hline
\hline
0 & 45.7\%, 11.8GeV, (7.41) & $>100$\%, -, - \\
\hline
1 & 26.5\%, 10.6GeV, (6.35) & 77.3\%, 11.6GeV, (6.72) \\
\hline
2 & 12.3\%, 9.58GeV, (5.57) & 40.7\%, 10.3GeV, (5.86) \\
\hline
4 & 3.71\%, 8.75GeV, (5.32) & 4.21\%, 8.77GeV, (5.02)\\
\hline
\hline
 \multicolumn{3}{|c|}{ $v_0=220$ km/s} \\
\hline
\hline
0 & 35.1\%, 9.97GeV, (7.87) & $>100$\%, -, - \\
\hline
1 & 20.4\%, 9.16GeV, (6.56) & 42.0\%, 9.59GeV, (7.09) \\
\hline
2 & 12.5\%, 8.37GeV, (6.30) & 14.3\%, 8.92GeV, (6.07) \\
\hline
4 & 4.96\%, 7.64GeV, (9.14) & 5.43\%, 7.76GeV, (7.27) \\
\hline
\hline
\multicolumn{3}{|c|}{ $v_0=270$ km/s} \\
\hline
\hline
0 & 40.9\%, 8.96GeV, (7.49) & 57.6\%, 9.64GeV, (8.03) \\
\hline
1 & 23.3\%, 8.15GeV, (6.99) &  27.9\%, 8.75GeV, (6.79)\\
\hline
2 & 13.9\%, 7.76GeV, (7.55) & 15.4\%, 7.96GeV, (7.19)\\
\hline
4 & 4.74\%, 6.98GeV, (16.2) & 5.90\%, 7.48GeV, (11.5)\\
\hline
  \end{tabular}
\caption{Minimum channeling fractions at $E_R = $ 3 keV allowed at 90\%
by all constraints for various
values of $v_0$ and form factors $F_{\rm DM}(q) \propto q^n$.  Also shown
are the dark matter mass at the minimum channeling fraction and, in
parentheses, the
$\chi^2_{\rm DAMA}$ for the best fit to the DAMA spectrum. In all cases,
$q_{\rm XENON} = 15.5$\%. }
\label{tab:fchans}
\end{table}

Finally, as usual, uncertainty in the halo model has a significant
impact on the model constraints.  We emphasize that it is
difficult to know precisely how to interpret model constraints
without better knowledge of the allowed range of dark matter halo velocity 
distributions.  The most significant effect of
different halo models tends to be whether the distribution is
tighter or broader, so one can qualitatively consider the effect of
different halo models by changing the average rotational velocity
parameter $v_0$. In Table \ref{tab:fchans}, we consider the effect of a 
modest change
in the halo distribution by considering Maxwellian distributions
with $v_0 = 170$ and $270$ km/s in addition to the standard $v_0 = $220 km/s.
We present there the minimum channeling fraction at $E_R = $ 3keV allowed
by demanding consistency at 90\% with all experiments.  Note that,
while a higher power of $q$ in the form factor decreases the required channeling
fraction, one may not continue indefinitely to higher powers since the best
fit to DAMA eventually worsens.  Furthermore, when the strongest constraint
is XENON, larger values of $v_0$ are favored, since the broader velocity
distribution pushes the DAMA favored region to lower masses. However,
when CDMS-Si is the strongest constraint, lower values of $v_0$ are favored,
since larger dark matter masses enhance the ratio of the reduced 
mass $\mu$ at iodine vs. at silicon.  

Figure \ref{fig:qandpow} presents the effect from different possibilities
for the details of channeling itself, for three halo models
($v_0 = 170, 220, 270 {\rm km/s}$) and XENON quenching factors
($q_{\rm XENON} = 15.5$\%, 14.3\%).  Since it is unlikely that 100\%
of the energy of channeled events gets deposited electronically, we 
consider lower values for this ``channeled quenching factor'' 
$q_{\rm chan}$ \footnote{In reality, even this parameterization is still a qualitative 
approximation, and one expects that at any recoil energy there is a
distribution of $q_{\rm chan}$'s.}.  For a standard WIMP coupling, relatively high 
values $q_{\rm chan} \gtrsim 90$\% or large channeling fractions $\gtrsim$ 30\%
are necessary. However, for a
$q^2$ form factor, even $q_{\rm chan} \approx $70\% may be sufficient with as 
little as 15\% of events channeled.  We also show the effect of
varying the energy profile of the channeling fraction, 
parameterized as $f_{\rm I}(E_R) \propto E_R^{-\alpha}$.  For a standard
WIMP coupling, the minimum channeling fraction required at $E_R = $3keV
depends significantly on $\alpha$, and the profile must be fairly flat
to allow $f_{\rm chan} < $50\%.  Again, the situation is much improved
with a form factor.  For $F_{\rm DM} \propto q^2$, the dependence on
$\alpha$ is much weaker, and relatively large $\alpha$'s still allow
$f_{\rm chan} \sim $10\%.

\section*{Acknowledgments}
We have benefited from discussions with D. Dauvergne.
BF is supported by DOE grant DE-FG02-01ER-40676, ALF
is supported by DOE grant DE-FG02-01ER-40676 and NSF CAREER grant
PHY-0645456, and EK is supported by DOE grant DE-FG02-01ER-40676,
NSF CAREER grant PHY-0645456, and an Alfred P. Sloan Fellowship.
BT is supported by the Leon Madansky Fellowship.

\end{document}